\documentclass{jpp}
\usepackage{graphicx}

\graphicspath{{./}{figures/}}

\usepackage[utf8]{inputenc}
\usepackage[T1]{fontenc}
\usepackage{amsmath}
\usepackage{amssymb}
\usepackage{dsfont}

\newcommand{\sgn}{\operatorname{sgn}}

\usepackage{unicode}
\newcommand{\csso}{\fontencoding{LECO}\selectfont\char215}
\newcommand{\iotaslash}{\iota\hspace*{-0.45em}\text{\csso}}

\newcommand{\vpar}{\ensuremath{v_{\parallel}}}
\newcommand{\vperp}{\ensuremath{v_{\perp}}}
\newcommand{\xpar}{\ensuremath{x_{\parallel}}}
\newcommand{\xperp}{\ensuremath{x_{\perp}}}

\providecommand\bnabla{\boldsymbol{\nabla}}
\providecommand\bkappa{\boldsymbol{\kappa}}

\providecommand\bnabla{\boldsymbol{\nabla}}

\def \al {\mbox{$\alpha$}}

\def \kperp {\mbox{$k_{\perp}$}}

\def \vth {\mbox{$v_{\mathrm{T}}$}}

\def \bhat {\mbox{$\mathbf{\hat{b}}$}}

\def \l {\mbox{$\ell$}}

\def \eps {\mbox{$\epsilon$}}

\renewcommand \o {\mbox{$\omega$}}

\def \osT {\mbox{$\omega_*^{\mathrm{T}}$}}

\def \ost {\mbox{$\widetilde{\omega}_*$}}

\def \od {\mbox{$\omega_d$}}

\def \odt {\mbox{$\widetilde{\omega}_d$}}

\def \Reff {\mbox{$R_{\mathrm{eff}}$}}

\shorttitle{An ITG-optimized quasi-helical stellarator}
\shortauthor{G. T. Roberg-Clark, P. Xanthopoulos, and  G. G. Plunk}

\title{Reduction of electrostatic turbulence in a quasi-helically symmetric stellarator via critical gradient optimization}

\author{G. T. Roberg-Clark\aff{1}
  \corresp{\email{gar@ipp.mpg.de}},
  P. Xanthopoulos \aff{1}
 \and G. G. Plunk \aff{1}}

\affiliation{\aff{1}Max-Planck-Institut F\"ur Plasmaphysik, D-17491, Greifswald, Germany}

\begin{document}

\maketitle

\begin{abstract}
We present a stellarator configuration optimized for a large threshold (``critical gradient'') for the onset of the ion temperature gradient (ITG) driven mode, which achieves the largest critical gradient we have seen in any stellarator. Above this threshold, gyrokinetic simulations show that the configuration has low turbulence levels over an experimentally relevant range of the drive strength. The applied optimization seeks to maximize the drift curvature, leading to enhanced local-shear stabilization of toroidal ITG modes, and the associated turbulence. These benefits are combined with excellent quasisymmetry, yielding low neoclassical transport and vanishingly small alpha particle losses.  Analysis of the resulting configuration suggests a trade-off between magnetohydrodynamic (MHD) and ITG stability.
\end{abstract}

\section{Introduction} \label{sec:intro}

The excitation of the ion temperature gradient (ITG) mode in magnetically confined fusion devices leads to turbulence that is responsible for energy losses, which reduce the plasma confinement needed for potential fusion performance. For instance, it has been argued that, during operation of the Wendelstein 7-X (W7-X) stellarator in electron heating scenarios, the ITG mode leads to so-called ``ion temperature clamping"  \citep{Beurskens2021a,Beurskens2022}, thus preventing the heating of ions in the plasma core above 2 keV. While resolving the location and extent of the fine-scale (ion Larmor radius) fluctuations is difficult in experiments, implementation of diagnostics using phase contrast imaging \citep{Baehner2021} and Doppler reflectometry \citep{Carralero2021} has helped to characterize the dependence of W7-X experimental temperature and density profiles on such fluctuations. 

In order to lessen the negative effects of the ITG mode, a possible strategy to follow involves the manipulation of the density and electron profiles, as implemented in W7-X using pellet injections \citep{Bozhenkov2020a,Pablant2020}. A stellarator magnetic field may also be shaped to reduce microturbulence losses during steady-state operation. In the case of electron-temperature gradient turbulence, it is argued that multiple field periods are stabilizing via reduction of the parallel connection length \citep{Plunk2019}, while trapped-electron mode turbulence can be ameliorated by requiring that the parallel adiabatic invariant J achieves its maximum on the magnetic axis \citep{Proll2012}, the so-called ``Maximum-J'' property \citep{Helander2012,Mackenbach2022}.

While most previous works have exploited magnetic field shaping in order to lower the rate of the ITG transport scaling ("stiffness"), see,  e.g., \citep{Mynick2010a,Xanthopoulos2014a,Hegna2018,Nunami2013,Jorge2021a,Stroteich2022}, the goal of the present work is to increase the linear onset of ITG modes ("critical gradient") \citep{Roberg-Clark2021}, since for configurations with stiff transport, it can determine the radial ion temperature profile \citep{Baumgaertel2013}. Even though the nature of the marginally unstable fluctuations may differ between low-magnetic-shear stellarators \citep{Zocco2018,Zocco2022,Bhattacharjee1983} [Floquet-like] and tokamaks \citep{Terry1982,Romanelli1989,Biglari1989a,Jenko2001a,Plunk2014a} [Toroidal-like], it seems that the size of the so-called ``drift curvature'', a factor appearing in the gyrokinetic equation, can be used to predict the critical gradient for most optimized stellarator configurations \citep{Roberg-Clark2022}. Here we leverage this predictive capability to generate a new quasi-helically-symmetric configuration with low neoclassical transport and large ITG critical gradient. It turns out that turbulent losses from ITG modes above this threshold are suppressed when compared with a well-known quasi-helically symmetric configuration, in particular near the critical gradient. We attribute this enhanced stability, at least in part, to the high onset gradient of localized, toroidal ITG modes, which are further damped by flux expansion and local shear effects.

The paper is structured as follows. In section 2 we define the linear gyrokinetic system used to analyze ITG mode properties. Section 3 describes the optimization procedure used to generate the new configuration, while Section 4 presents results of the optimization, along with a description of the local shear effect. We conclude with an outlook on the apparent competition between ITG and MHD stability in section 5.

\section{Linear gyrokinetic equation}

Following \citep{Plunk2014a}, we use the standard gyrokinetic system of equations \citep{Brizard2007} to describe electrostatic fluctuations destabilized along a thin flux tube tracing a magnetic field line. The ballooning transform \citep{Dewar1983a} and twisted slicing representation \citep{Roberts1965} are used to separate out the fast perpendicular (to the magnetic field) scale from the slow parallel scale. The magnetic field representation in field following (Clebsch) representation reads, $\mathbf{B}=\bnabla \psi \times \bnabla \al$, where $\psi$ is a toroidal flux surface label and $\alpha=\vartheta - \iota\phi$ labels the magnetic field line on the surface, with $q$ the safety factor, $\vartheta$ the poloidal angle, and $\phi$ the toroidal angle. The perpendicular wave vector is then expressed as $\mathbf{k_{\perp}} = k_{\alpha} \bnabla \alpha + k_{\psi} \bnabla \psi$, where $k_{\alpha}$ and $k_{\psi}$ are constants, so the variation of $\mathbf{k_{\perp}}(l)$ stems from that of the geometric quantities $\bnabla \alpha$ and $\bnabla \psi$, with $\l$ the field-line-following (arc length) coordinate.

We assume Boltzmann-distributed (adiabatic) electrons, thus solving for the perturbed ion distribution $g_{i}(\vpar,\vperp,\l,t)$, defined to be the non-adiabatic part of $\delta f_{i}$ ($\delta f_{i}=f_{i}-f_{i0})$ with $f_{i}$ the ion distribution function and $f_{i0}$ a Maxwellian. The electrostatic potential is $\phi(\mathbf{\l})$, and $\vpar$ and $\vperp$ are the particle velocities parallel and perpendicular to the magnetic field, respectively. 

The linear gyrokinetic equation for the ions is written
\begin{equation}
i\vpar \frac{\partial g}{\partial \ell} + (\o - \odt)g = \varphi J_0(\o - \ost)f_0\label{gk-eqn}
\end{equation}\label{eqn:lingyro}

\noindent with the following definitions: $J_0 = J_0(k_{\perp}v_{\perp}/\Omega) = J_0(k_{\perp}\rho\sqrt{2}v_{\perp}/\vth)$; the ion thermal velocity is $\vth = \sqrt{2T/m}$ and the thermal ion Larmor radius is $\rho = \vth/(\Omega\sqrt{2})$; $n$ and $T$ are the background ion density and temperature; $q$ is the ion charge; $\varphi = q\phi/T$ is the normalized electrostatic potential; $\Omega=q B/m$ is the cyclotron frequency, with $B=|\mathbf{B}|$ the magnetic field strength.  Assuming Boltzmann electrons, the quasineutrality condition is

\begin{equation}
\int d^3{\bf v} J_0 g = n(1 + \tau) \varphi,\label{qn-eqn}
\end{equation}\label{eqn:poisson}

\noindent where $\tau = T/(ZT_e)$ with the charge ratio defined as $Z = q/q_e$. The equilibrium distribution is the Maxwellian

\begin{equation}
f_0 = \frac{n}{(\vth^2\pi)^{3/2}}\exp(-v^2/\vth^2),
\end{equation}

\noindent and we introduce the velocity-dependent diamagnetic frequency

\begin{equation}
\ost = \osT \left[\frac{v^2}{\vth^2} - \frac{3}{2}\right]
\end{equation}

\noindent where we neglect background density variation and define $\osT = (Tk_{\alpha}/q)d\ln T/d\psi$.  The magnetic drift frequency is $\odt = {\bf v}_d\cdot{\bf k}_{\perp}$ and the magnetic drift velocity is ${\bf v}_d = \hat{\bf b}\times((v_{\perp}^2/2)\bnabla \ln B  + \vpar^2\bkappa)/\Omega$, where $\bkappa = \hat{\bf b}\cdot\bnabla\hat{\bf b}$. We take $\bnabla \ln B = \bkappa$, the zero $\beta$ approximation, for simplicity. We then let

\begin{equation} \label{eqn:omegadrift}
\odt = \frac{\mathbf{k_{\perp}} \cdot (\bhat \times \bkappa)v^{2}_{T}}{\Omega} \left[\frac{\vpar^2}{\vth^2} + \frac{\vperp^2}{2\vth^2}\right] = \od(\l) \left[\frac{\vpar^2}{\vth^2} + \frac{\vperp^2}{2\vth^2}\right],
\end{equation}

\noindent where the velocity-independent drift frequency $\od(\ell)$ generally varies along the field line. Positive values of $\od$ correspond to ``bad'' curvature, i.e. are destabilizing for ITG modes, assuming $\osT$ is negative. Assuming $k_{\psi}=0$ for simplicity, we then define the drift curvature $K_{d}$ by writing

\begin{equation} \label{eqn:kd}
    \od(\ell) \propto K_{d}(\ell) \equiv a^2{\bnabla}\alpha \cdot \bhat \times \bkappa,
\end{equation}
where $K_{d}$ contains the purely geometric variation of the drift frequency and $a$ is the minor radius of the flux surface at the edge. For the purpose of analyzing gyrokinetic simulation results we define the metrics $g^{yy}=a^{2}s_{0}/(q_{0}^{2})( \bnabla \alpha)^{2}$, with $s=\psi/\psi_{edge}$ the toroidal flux normalized to its value at the last closed flux surface ($q_{0}$ is the safety factor on a particular surface $s=s_{0}$), and $g^{xx}=a^{2}/(4s_{0})(\bnabla s)^{2}$. Finally, we define the poloidal wavenumber $k_{y}=(q_{0}/\sqrt{s_{0}})k_{\alpha}$.

\subsection{Averaging effect in the gyrokinetic equation}

An integral equation can be derived from Eqns.~\ref{gk-eqn}-\ref{qn-eqn} assuming ``outgoing'' boundary conditions $g(\vpar > 0, \ell = -\infty) = g(\vpar < 0, \ell = \infty) = 0$, consistent with ballooning modes that decay as $|\ell| \rightarrow \infty$ \citep{Connor1980,Romanelli1989}.  To enforce these conditions we assume the system has non-zero global shear, $d\iotaslash/d\psi \neq 0$, though it is allowed to be small. One then obtains \citep{Plunk2014a}

\begin{align}
(1+ \tau)\varphi(\ell) = \frac{-2i}{\vth\sqrt{\pi}}\int_{0}^{\infty} \frac{d\xpar}{\xpar} \int_{0}^{\infty}  d\xperp \xperp (\o - \ost) J_0 \nonumber \\ 
\times \int_{-\infty}^{\infty} d\ell^{\prime} J_0^{\prime} \exp(-x^2 + i \sgn(\ell-\ell^{\prime})M(\ell^{\prime}, \ell))\varphi(\ell^{\prime}),\label{eqn:ballooning-disp}
\end{align}

\noindent where $\xperp = \vperp/\vth$ and $\xpar = \vpar/\vth$, $\sgn$ gives the sign of its argument, $J_0 = J_0 \left(\sqrt{2b(\ell)}\xperp \right)$, $J_0^\prime = J_0 \left(\sqrt{2b(\ell^\prime)}\xperp \right)$, and $b(\ell)=\rho^{2}k_{\perp}^{2}(\ell)$. The physics of the drift resonance is contained in the factor

\begin{equation}\label{eqn:mbar}
M(\ell^{\prime}, \ell) = \int_{\ell^{\prime}}^{\ell} \frac{\o - \odt(\ell^{\prime\prime})}{\vth \xpar } d\ell^{\prime\prime}.
\end{equation}

The kernel M contains an averaging over $\ell^{\prime\prime}$ which suggests the use of a smoothed curvature profile, where small-scale ripples are averaged out and a ``coarse-grained'' drift curvature can then be substituted into (\ref{eqn:ballooning-disp}) (see \cite{Roberg-Clark2022} for further discussion).

We neglect particle trapping so $\xpar$ and $\xperp$ do not depend on $\ell$. Most evidence indicates that the ITG mode uniformly responds to changes in the parameters $\tau$ and $\mathrm{d}n/\mathrm{d}\psi$ and is stabilized by increasing either of them, except for a relatively small region of parameter space where positive density gradients can destabilize the mode. For simplicity we thus set $\tau=1$ and $\bnabla n=0$.

\subsection{Estimating the ITG mode critical gradient}

As found in Roberg-Clark et al. (2022), the geometric dependence of the linear ITG critical gradient can be estimated in a simple way. The drift curvature profile (eqn. \ref{eqn:kd}) on a particular magnetic field line (at radial location $s = s_{0}$ and field line $\alpha=\alpha_{0}$) is fitted with quadratic curves in regions of bad curvature, which we refer to as ``drift wells''. The fitting acts an effective coarse-graining of the geometry, producing a smoothed curvature profile as mentioned in the preceding section. Using the fitted profile, a value of the predicted critical gradient is produced, corresponding to the drift well with the smallest critical gradient. The formula reads:
\begin{equation}\label{eqn:critgrad}
 F_{crit}=\frac{a}{L_{T,crit}}=2.66\left(\frac{a}{\Reff}\right)   
\end{equation}
with $a/\Reff$ the peak value of the drift curvature within the drift well. Here we have ignored the parallel stabilizing term included in Roberg-Clark et al. (2022) as we intend to produce a configuration with a large critical gradient solely by large ``drift curvature''; see the later discussion of geometric interpretation in Section \ref{sec:ITG}.

\section{Optimization Method}

We use the SIMSOPT software framework \citep{Landreman2021} to generate a vacuum stellarator configuration with quasi-helical (QH) symmetry and a significant linear ITG critical gradient as a result of drift curvature. The stellarator magnetic field is described by a boundary surface given in the Fourier representation 
\begin{align} \label{eqn:Fourier}
    R(\vartheta,\phi)=\sum_{m,n} R_{m,n} \cos(m\vartheta-4 n\phi) \\
    Z(\vartheta,\phi)=\sum_{m,n} Z_{m,n} \sin(m\vartheta-4 n\phi),
\end{align}
where we have assumed stellarator symmetry and have set the number of field periods to $n_{fp}=4$. Global vacuum solutions are constructed at each iteration by running the VMEC \citep{Hirshman1983a} code, which solves the MHD equations using an energy-minimizing principle. 

Optimization proceeds by treating the Fourier coefficients in equation (\ref{eqn:Fourier}) as parameters and varying them in order to find a least-squares minimization (using a trust region) of the specified objective function, which reads
\begin{equation} \label{eqn:objfunc}
    f = f_{QS} + (A-4.10)^{2} + \left(F_{crit} - 2.00 \right)^2,
\end{equation}
where 
\begin{equation}
    f_{QS} = \sum_{s_{j}}\left \langle \left( \frac{1}{B^{3}}[(-1-\iota)\mathbf{B}\times \bnabla B \cdot \bnabla \psi - G\mathbf{B} \cdot \bnabla B] \right)^{2} \right \rangle
\end{equation}
is the quasisymmetry residual taken from \citep{Landreman2022}, with $A=R/a$ the aspect ratio output by VMEC, and $a/L_{T,crit}$ as defined in eqn. (\ref{eqn:critgrad}) with the field line $(s_{0}=0.5$, $\alpha_{0}=0)$ chosen. Here $G(\psi)$ is $\mu_{0}/(2\pi)$ times the poloidal current outside the surface and we have set the toroidal current to zero, while $\langle \cdot \rangle$ denotes a flux surface average over the individual surfaces $s=s_{j}$. We choose $B=B(\theta+4\phi_{B})$, with $\theta$ and $\phi_{B}$ the Boozer poloidal and toroidal angles, corresponding to quasi-helical symmetry with a negative axis helicity. The input equilibrium for the optimization is a "warm start" example file included in SIMSOPT with approximate quasi-helical symmetry, $n_{fp}=4$ and aspect ratio $A=R/a=6$, with $R$ the major radius and $a$ the minor radius. The target $A=4.10$ was chosen to enhance the drift curvature $K_{d}$, while we speculate $a/L_{T,crit}=2.00$ to be roughly the maximum achievable linear critical gradient in the absence of shear stabilization \citep{Roberg-Clark2022}. We calculate the residual $f_{QS}$ on the surfaces $s=[0,0.1,0.2,0.3,0.4,0.5]$, thus leaving the region $0.5<s<1.0$ free. We make this choice knowing that requiring a ``precise'' degree of quasisymmetry in the entire volume often has the effect of reducing the global magnetic shear in the configuration to vanishingly small values \citep{Landreman2022}. Similarly to \cite{Landreman2022}, the number of Fourier coefficient parameters are increased in a series of four steps, with the toroidal and poloidal mode numbers in the VMEC calculation chosen to be $m_{pol}=n_{tor}=[3,5,6,6]$. The final step is a refinement to see if a better local minimum can be found at the same resolution of six modes.

\section{Results}

\begin{figure}
    \centering
    \includegraphics[scale=.32]{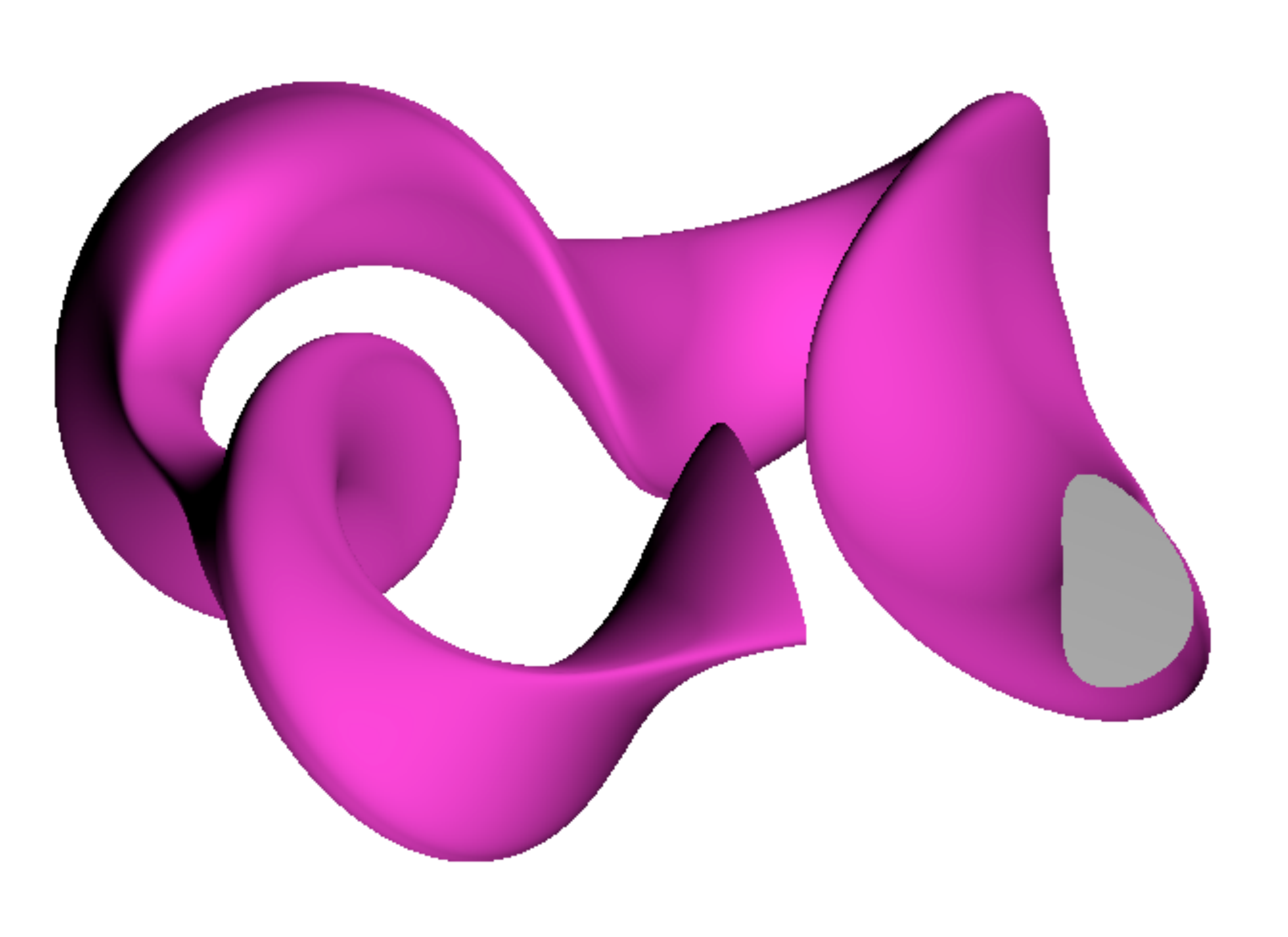}
    \caption{The boundary surface of HSK.}
    \label{fig:N4A4}
\end{figure}

\begin{figure}
    \centering
    \includegraphics[scale=.25]{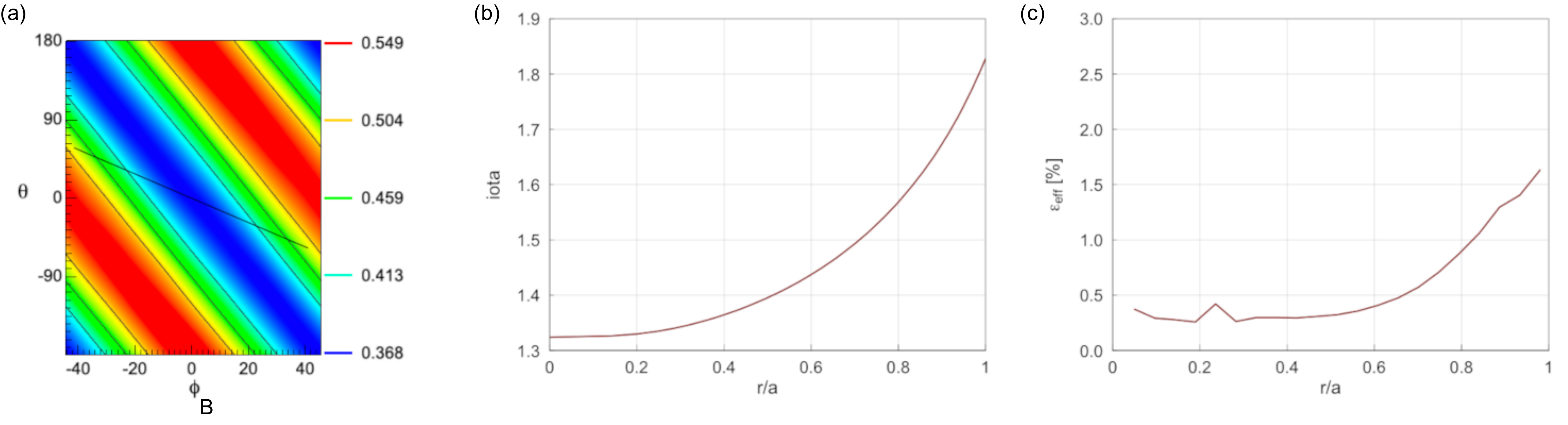}
    \caption{Properties of HSK. (a) Contours of $B$ with the trajectory of a field line (black curve near the center) overlaid in Boozer coordinates. (b) Rotational transform as a function of radius. (c) Neoclassical transport coefficient $\eps_{eff}$ as a function of radius.}
    \label{fig:N4A4stuff}
\end{figure}

\begin{figure}
    \centering
    \includegraphics[scale=.32]{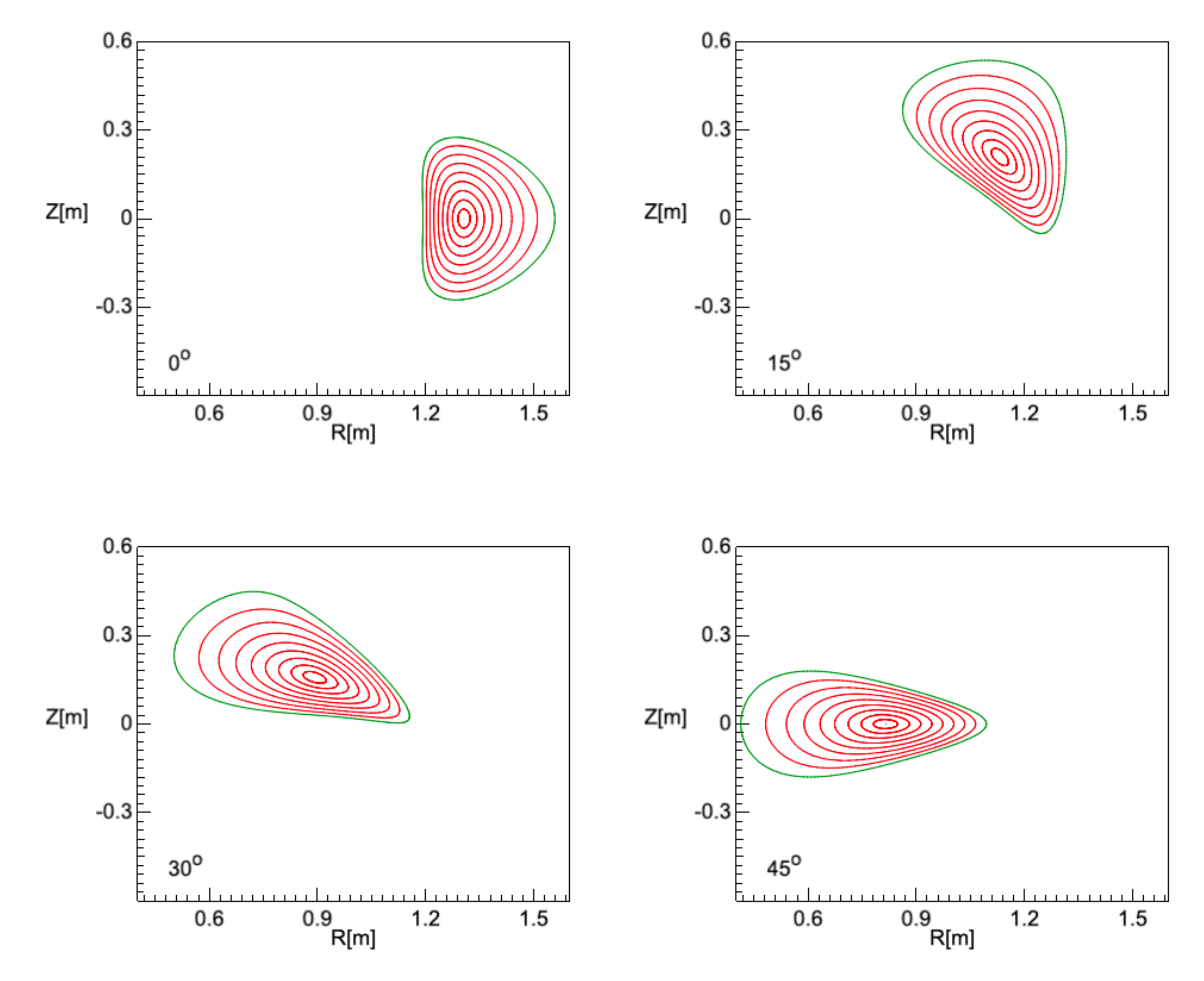}
    \caption{Surface cuts of HSK taken at constant toroidal angle, $\phi=\text{const}$.}
    \label{fig:cuts}
\end{figure}

Figure \ref{fig:N4A4} shows a surface plot of the outermost flux surface of the resulting optimized configuration, which we dub ``HSK'' (Helically-Symmetric Kompakt stellarator). The contours of $B$ are plotted in the Boozer angle plane at $s=0.25$ in Fig. \ref{fig:N4A4stuff}(a) indicating the quasi-helical symmetry. The highly compact configuration has roughly triangular cross sections (cuts at constant toroidal angle) [Fig. \ref{fig:cuts}] and noticeably lacks a ``bean-shaped'' cross section at toroidal angle $\phi=0$, which most optimized stellarator configurations possess. The spatial separation between the flux surfaces is substantial, indicating low surface compression, while the lack of an indentation on the inboard side at $\phi=0$ appears to be a result of not optimizing for a vacuum magnetic well. The configuration instead has a sizable magnetic hill, i.e. $V^{\prime\prime}(\psi)$, the second derivative of the surface volume, is positive at all radii. Heuristically speaking, this result is expected from configurations with average bad curvature, i.e. the average value of the drift curvature over the entire surface is positive. HSK has a rotational transform varying from $\iota = 1.3$ to $1.8$ (demonstrating significant global magnetic shear) [Fig. \ref{fig:N4A4stuff}(b)] and values of the neoclassical transport coefficient $\eps_{eff}$ \citep{Nemov1999} ranging from $0.4\%$ on axis to $1.6\%$ at the last closed flux surface [Fig. \ref{fig:N4A4stuff}(c)], implying excellent neoclassical confinement. Using a symplectic integrator for guiding center trajectories \citep{Albert2020} in this configuration but rescaled to the same $B$ and minor radius as the ARIES-CS reactor \citep{Mau2008}, we find no collisionless alpha particle losses after 10 $ms$ for particles launched from inner radii, suggesting excellent quasisymmetry (figure not shown).

\begin{figure}
    \centering
    \includegraphics[scale=.65]{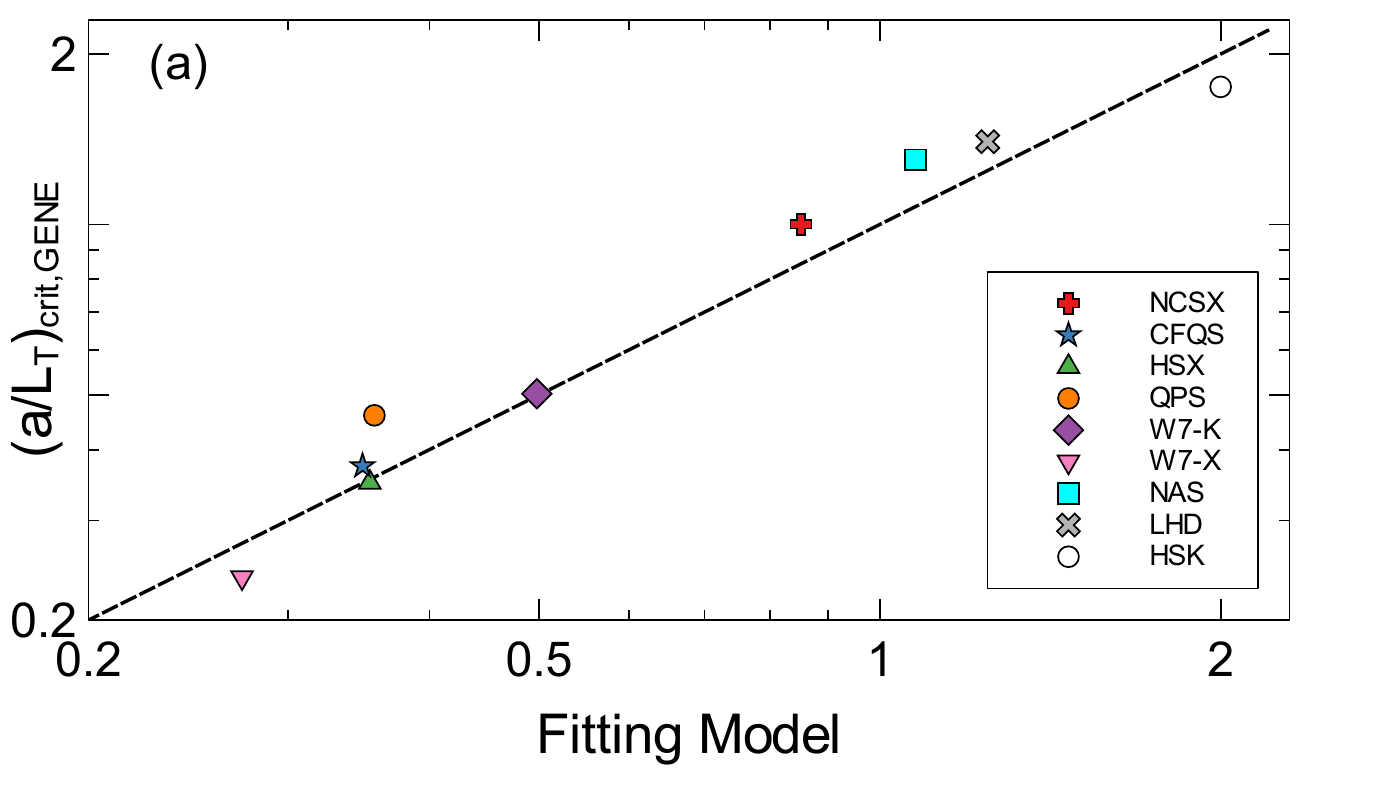}
    \caption{Simulation of critical gradients versus model predictions on a log-log scale for flux tubes taken from several stellarator geometries, adapted from figure (1) in \citep{Roberg-Clark2022}, with the data point for HSK added. The field line $(s_{0}=0.5,\alpha_{0}=0)$ was chosen for all configurations.}
    \label{fig:critgrad}
\end{figure}

Linear gyrokinetic simulations in local flux tube geometry using the GENE code \citep{Jenko2000a} are performed to determine the critical gradient in the same manner as in \cite{Roberg-Clark2021,Roberg-Clark2022}, i.e. by reducing the applied temperature gradient until a single, marginally unstable mode remains. An unusually large resolution in $\mu$ (i.e. $\vperp$ space) of $n_{\mu}=32$ was required for numerical convergence of the critical gradient, likely because of the extreme values of curvature near the outboard midplane that imply significant linear phase mixing in $\vperp$ through the $\bnabla B$ drift term in eqn. (\ref{eqn:omegadrift}). The linear ITG critical gradient at $(s_{0}=0.5, \alpha_{0}=0)$ [$k_{y}\rho=0.3$], while not reaching $2$ as predicted by the fitting model (eqn. \ref{eqn:critgrad}), nonetheless attains the value of $a/L_{T,crit}=1.75$, the highest we have seen in any stellarator (Fig. \ref{fig:critgrad}). Thus the simultaneous optimization for quasi-helical symmetry, aspect ratio, and ITG linear critical gradient was successful. 

One might reasonably expect this particular ITG optimization strategy (increasing bad curvature) to heavily exacerbate linear growth rates, and thus nonlinear transport, of ITG modes above the linear critical gradient, implying a tradeoff between linear and nonlinear stability. Our results in the next section, however, oppose this intuition, revealing significant ITG stability above the critical gradient.

\subsection{Mechanisms of ITG turbulence suppression}\label{sec:ITG}

\begin{figure}
    \centering
    \includegraphics[scale=0.4]{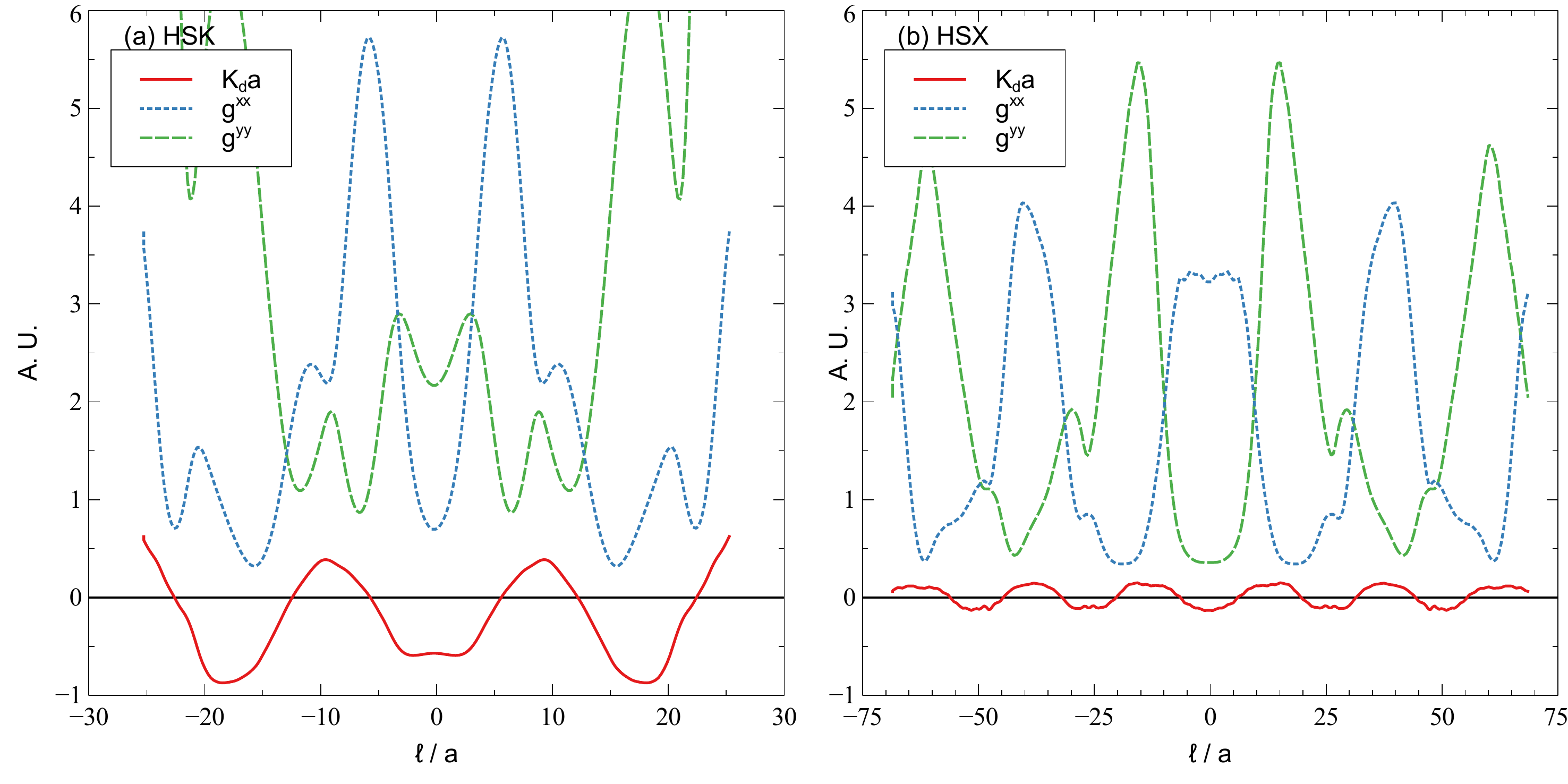}
    \caption{Metrics (defined near the end of section 2) as a function of the field-line-following coordinate $\ell$ at the flux tube location $(s_{0}=0.5,\alpha_{0}=0)$, where HSK was optimized for a high critical gradient. The flux tubes were constructed with two poloidal turns. The vertical axis is in dimensionless units, such that the metrics for each configuration can be compared directly. (a) HSK. (b) HSX.}
    \label{fig:metrics}
\end{figure}

One means by which a stellarator configuration can attain large values of the drift curvature $K_{d}$ (eqn. \ref{eqn:kd}) is by compression of the $\alpha$ coordinate, {\em i.e.} large $|\bnabla \alpha|$. This is indeed observed for HSK, in the metrics following the magnetic field line at $(s_{0}=0.5,\alpha_{0}=0)$, where we plot $g^{yy} \propto g^{\alpha\alpha}=a^{2}|\bnabla \al|^{2}$ as a function of arc length $\ell$, which attains a value on the outboard midplane ($\ell=0$) larger than $2$ [Fig. \ref{fig:metrics}(a)]. Note that this location at the outboard midplane corresponds to $\phi=0$ in figure \ref{fig:N4A4}(b). For comparison, we also plot the metrics along the same field line in the the HSX stellarator \citep{Talmadge2008}, another QH optimized configuration [Fig. \ref{fig:metrics}(b)]. 

In a simple limit, when $\bnabla\alpha\cdot\bnabla \psi \approx 0$, and assuming weak variation of the overall magnetic field strength $B = |\bnabla \al \times \bnabla \psi|$, the $\alpha$-compression is directly linked to reduction of $g^{xx}\propto g^{\psi\psi}$ ({\em i.e.} $g^{xx} \sim 1/g^{yy}$).  In such cases, reduced instability growths and turbulent transport can be intuitively explained via expansion of flux surfaces, resulting in an effective suppression of the applied temperature gradient, since the physical temperature gradient $|\bnabla T|=dT/d\psi |\bnabla \psi|$ scales with $|\bnabla \psi|$ \citep{Plunk2022,Helander2021,Stroteich2022,Angelino2009}.

Typically in a stellarator the vector $\bnabla \alpha$ also develops a substantial component in the direction parallel to $\bnabla\psi$, attributed to so-called ``local shear'' of the magnetic field lines \citep{Helander2014}.  This effect, distinct from flux expansion, is associated with a stabilization effect \citep{Waltz1993} due to $\kperp \rho_i$ increasing, {\em i.e.} due a finite Larmor radius (FLR).  Global shear may further amplify this damping through secular increase of $g^{yy}$ along the field line, although we suspect this is not the dominant effect in the case of HSK.  We see that both of these effects are at play with HSK, but note for the case of HSX that the pattern near $\l=0$ in the metrics is effectively inverted, i.e. $|K_{da}|<<1$, $g^{yy}<1$ and $g^{xx}>1$, which is typical of most optimized stellarator configurations.

\begin{figure}
    \centering
    \includegraphics[scale=0.22]{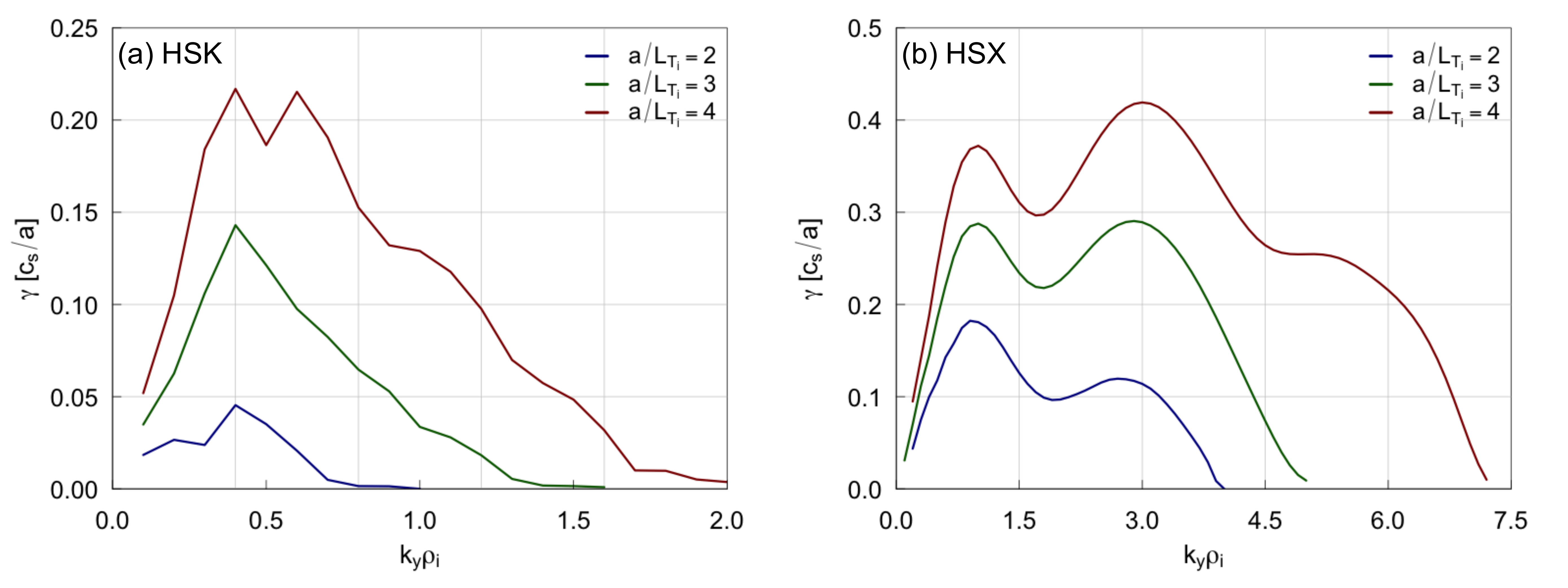}
    \caption{ITG mode linear growth rates as a function of $k_{y} \rho_{i}$ for HSK (a) and HSX (b) for the temperature gradients $a/L_{T}=2,3$, and $4$, at the flux tube location $(s_{0}=0.25,\alpha_{0}=0)$.}
    \label{fig:ITGlin}
\end{figure}

We now carry out linear gyrokinetic simulations comparing HSK with HSX using the GENE code, but this time above marginal stability. We choose the flux tube $(s_{0}=0.25,\alpha_{0}=0)$ to check that the benefits of HSK are not confined to the field line optimized for ITG stability at $(s_{0}=0.5,\alpha_{0}=0)$. Figure \ref{fig:ITGlin}(a) shows a strikingly narrow range of unstable wavenumbers for HSK in $k_{y}\rho$ at the gradient $a/L_{T}=2$. Much of the analysis of electrostatic modes in QH stellarators has focused on damped or subdominant eigenmodes \citep{Faber2018,Terry2006,Sugama1999a}, which can form the basis for nonlinear saturation of ITG turbulence \citep{Pueschel2016a,Hegna2018,McKinney2019a}. We interpret the narrow linear growth rate spectrum of HSK at $a/L_{T}=2$ to mean that most of the linear eigenmode spectrum is stable at this gradient. It is worth mentioning that HSX sometimes has a favourable scaling of heat flux at these temperature gradients compared to other configurations \citep{McKinney2019a,Plunk2017}, despite larger linear growth rates. HSX was also not optimized for reduced turbulent transport, so one should not expect it to outperform turbulence-optimized configurations. 

\begin{figure}
    \centering
    \includegraphics[scale=0.23]{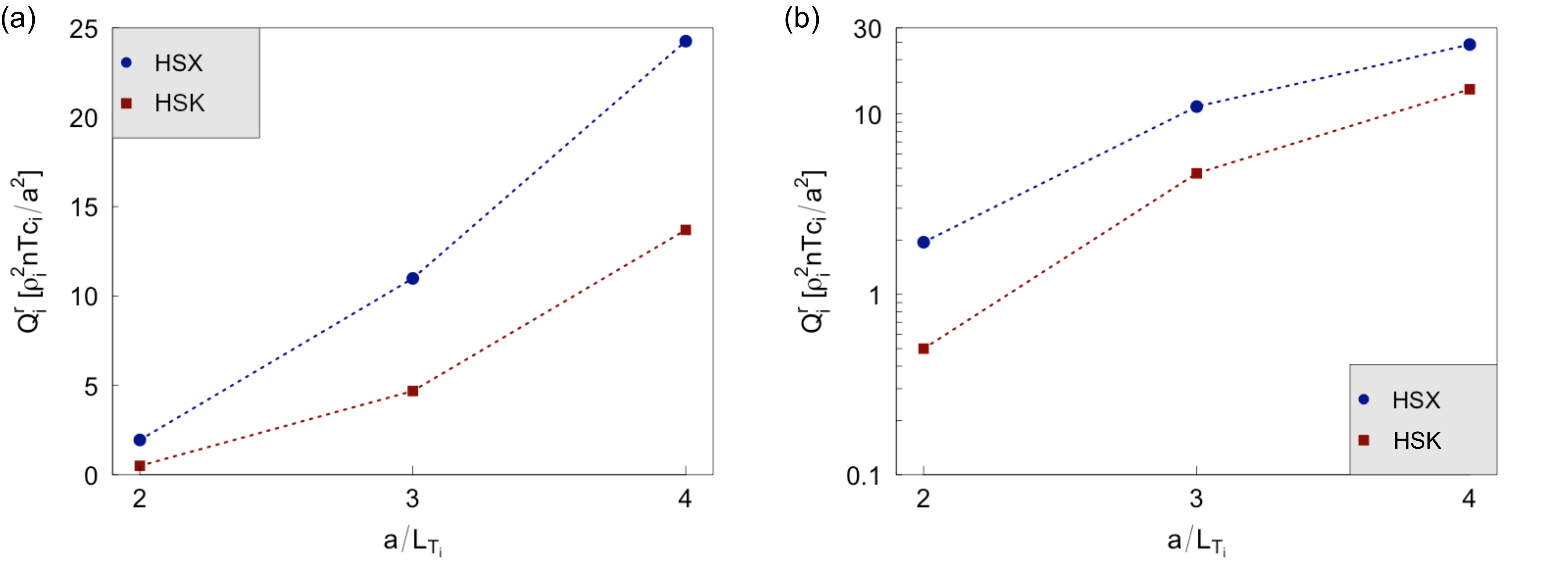}
    \caption{(a) Nonlinear heat fluxes computed by GENE at the flux tube $(s_{0}=0.25,\alpha_{0}=0)$ for HSK and HSX, varying the applied temperature gradient. (b) The same plot as (a) but with the vertical axis on a log scale.}
    \label{fig:ITGnlin}
\end{figure}

A nonlinear gyrokinetic analysis with GENE using the same flux tubes shows that the narrow and small linear growth rates in HSK are complemented by a stark reduction in heat fluxes, a truer figure of merit for confinement. In comparison with HSX, we find heat fluxes are smaller by roughly a factor of $4$ at $a/L_{T}=2$ (Fig. \ref{fig:ITGnlin}(b)). While the reduction is not as strong at higher gradients, the curve for HSK remains below that of HSX for the full range of gradients studied. The effect of the optimization is most potent near the critical gradient, $a/L_{T} \simeq 2 - 3$. This range of gradients is likely the most relevant to experiments, as seen in data for \citep{Beurskens2021a} and in global gyrokinetic simulations of \citep{Navarro2022} W7-X. For the same heating power, we can thus expect HSK to have a significantly larger temperature gradient compared to HSX, leading to improved confinement and higher core ion temperatures. Furthermore, HSK has the advantage of a reduced aspect ratio, which implies a favorable confinement time compared to larger aspect ratio devices with the same transport coefficients. Note that the turbulence-optimized WISTELL-C QH configuration presented in \cite{Hegna2022} is likely compared with HSX at the radial location $s=0.5$, achieving a factor of 2 reduction in heat fluxes at $a/L_{T}=4$. The same reduction factor for $a/L_{T}=4$ at $s_{0}=0.25$ is found when comparing HSK and HSX [Fig. \ref{fig:ITGnlin}(b)].

While bad curvature has markedly increased in HSK relative to HSX, stiffness of ITG transport has not worsened for the range $2 \leq a/L_{T} \leq 4$. We attribute this surprising result to the fact that the onset of localized, toroidal ITG modes \citep{Zocco2018} appears to have been increased by the combined effect of bad curvature and parallel stabilization by FLR damping, pushing it beyond the critical gradient $a/L_T,crit=1.75$. It has long been suspected that these modes, rather than extended Floquet or slab-like modes, cause the worst ITG transport \citep{Jenko2002a,Zocco2022}. Such an increase in the toroidal threshold could be modelled by an additional stabilization term in the expression (\ref{eqn:critgrad}) \citep{Jenko2001a} and will be explored in future work.

\section{Balancing MHD and ITG stability}

HSK is MHD unstable for reasonably finite values of $\beta$ so the comparison to a reactor with finite plasma pressure is theoretical. However, it is revealing to see the extent to which ITG turbulent losses can be quenched by abandoning MHD stability in the optimization. This trade-off was hinted at in previous reactor studies carried out for QH stellarators \citep{Bader2020,Bader2019}. We can gain some intuition for the lack of MHD stability in HSK by comparing to the standard ``bean-shaped'' cross section in W7-X, which has an indentation on the inboard, and vertical, compressed surfaces on the outboard. The indentation aids the formation of a vacuum magnetic well by strongly reducing the volumetric expansion of the surfaces such that $V^{\prime\prime}(\psi)<0$ \citep{Cooper1992}. The outboard of the bean section has a small drift curvature, a result of large values of $g^{\psi\psi}$ as well as a small geometric curvature [see discussion of the QIPC stellarator \citep{Subbotin2006,Beidler2011}, which has a similar bean-shaped section to that of W7-X]. Quantitatively, we also know from near-axis theory that the magnitude of the drift curvature contributes to the formation of an unstable magnetic hill \citep{Landreman2020}, and that the expression for Mercier stability (governing large-n ballooning modes near rational surfaces) in general contains destabilizing terms proportional to $1/|\bnabla \psi|$ \citep{Landreman2020}. Thus the bean cross-section, possessing large values of $|\bnabla \psi|$, a small geometric curvature, and $V^{\prime\prime}(\psi)<0$, is generally stable to MHD modes.

In contrast, HSK lacks an indentation and has expanded surfaces on the outboard side, leading to a magnetic hill and significant instability with respect to the Mercier criterion. However, it is plausible that stability of a configuration like HSK could be enhanced through the addition of an indentation and a weakening of the drift curvature on the outboard side. Optimization studies exploring this compromise between MHD and ITG stability are currently underway.

\subsection{Discussion}

Our optimization strategy for stellarators seems to have simple geometric consequences. The near-ubiquitous ``bean-shape'' cross section, which is thought to impart significant MHD stability and is often the site of the most detrimental ITG turbulence, can be modified to acquire a more triangular shape with a point at the outboard midplane via increasing the gradient of the binormal coordinate, $|\bnabla \alpha|$. The resulting increase in ``bad'' curvature leads to improved linear ITG mode critical gradients, while heat transport at gradients near this threshold is significantly reduced. The feared trade-off between large critical gradients and stiffness of the transport above those thresholds appears not to be a true impediment, as also hinted at by the compact W7-K configuration \citep{Roberg-Clark2022}. Rather than reducing the magnitudes of both drift curvature (eqn. \ref{eqn:kd}) and flux surface compression \citep{Mynick2010a,Xanthopoulos2014a}, one can increase drift curvature and local shear while improving both the ITG critical gradient and the stiffness of near-marginal transport. The reason appears to be that the critical gradient of localized, toroidal ITG modes is also increased, shielding the configuration from the most detrimental transport losses.

We note that the current work has not taken micro-turbulence effects into account such as kinetic electron physics of ITG modes \citep{Helander2015,Proll2022}, non-zero density gradients \citep{Thienpondt2022}, trapped-electron mode turbulence \citep{Faber2015,Mackenbach2022}, or finite beta effects, which are generally stabilizing \citep{Pueschel2008a,Zocco2015}. However, our goal here is to study a worst-case scenario for ITG modes and how it can be improved from purely geometric considerations, while also ensuring good quasisymmetry. Our optimization strategy and gyrokinetic analysis are local, specific to certain flux tubes, and future work will address ITG stability on the entire surface, taking more locations into account to ensure the optimization succeeds globally.

The most salient compromise to emerge from this work is that between ITG and MHD stability, and future designs will likely have to prioritize which type of stability is most important. We note that a similar optimization trade-off appears to exist between quasisymmetry and MHD stability measures such as the vacuum magnetic well \citep{Landreman2022,Landreman2022a}. There are promising signs that Mercier stability does not have to be rigidly adhered to, e.g. in global gyrokinetic simulations of kinetic ballooning modes \citep{Mishchenko2022}, in order for heat fluxes at finite beta to match those of traditional MHD-optimized stellarators such as W7-X. We plan to study the stability of HSK to kinetic ballooning modes using this approach to see how detrimental electromagnetic turbulence can be when Mercier stability is strongly violated. Furthermore, there is experimental evidence from the LHD heliotron \citep{Fujiwara2001} and stellarators such as TJ-II \citep{deAguilera2015} and W7-AS \citep{Geiger2004,Weller2006} that Mercier-unstable and/or magnetic hill configurations can operate at relatively large $\beta$ values with no serious loss of confinement, with the caveat that the LHD results may be restricted to low density operational regimes. A wide range of stellarator configurations, straddling the line between rigorous MHD stability and strongly suppressed ITG turbulence, thus appears realizable.

\textit{Acknowledgments}. The authors thank C. N\"uhrenberg for checking MHD stability of the HSK configuration, R. Jorge for providing the NEAT code to track alpha particle losses, and M. Landreman and B. Medasani for help with the use of SIMSOPT. We thank P. Helander, A. Zocco, and M. Landreman for contributing useful ideas to this work. This research was supported by a grant from the Simons Foundation (No. 560651, G. T. R.-C.). Computing resources at the Cobra cluster at IPP Garching and the Marconi Cluster were used to perform the simulations.  This work has been carried out within the framework of the EUROfusion Consortium, funded by the European Union via the Euratom Research and Training Programme (Grant Agreement No 101052200 — EUROfusion). Views and opinions expressed are however those of the author(s) only and do not necessarily reflect those of the European Union or the European Commission. Neither the European Union nor the European Commission can be held responsible for them.

\bibliographystyle{jpp}
\bibliography{library.bib}

\end{document}